\documentclass{aa}
\usepackage{graphicx}
\usepackage{amsmath}
\usepackage{natbib}
\bibpunct{(}{)}{;}{a}{}{,}
\def\dt{\mathrm}
\def\ud{\mathrm{d}}

\begin{document}
\title{Thermodynamic evolution of the cosmological baryonic gas : II.
  Galaxy formation}

   \author{J.-M. Alimi
          \inst{1}
          \and
          S. Courty
          \inst{1,2}}

   \offprints{S. Courty}

   \institute{Laboratoire de l'Univers et de ses Th\'eories, CNRS UMR 8102,
     Observatoire de Paris-Meudon, 5 place Jules Janssen, 92195 Meudon, France\\
     \email{jean-michel.alimi@obspm.fr} \and
     Present address: Science Institute, University of Iceland, Dunhagi 3, 107 Reykjavik, Iceland \\
     \email{courty@raunvis.hi.is} }

   \date{Received date / accepted date}
   
   \abstract{The problem of galaxy formation and its dependence on
     thermodynamic properties is addressed by using Eulerian
     hydrodynamic numerical simulations of large scale structure
     formation. Global galaxy properties are explored in simulations
     including gravitation, shock heating and cooling processes, and
     following self-consistently the chemical evolution of a
     primordial composition hydrogen-helium plasma without assuming
     collisional ionization equilibrium. The galaxy formation model is
     mainly based on the identification of converging dense cold gas
     regions. We show that the evolution at low redshift of the
     observed cosmic star formation rate density is reproduced, and
     that the galaxy-like object mass function is dominated by
     low-mass objects. The galaxy mass functions are well described by
     a two power-law Schechter function whose parameters are in good
     agreement with observational fits of the galaxy luminosity
     function. The high-mass end of the galaxy mass function includes
     objects formed at early epochs and residing in high-mass dark
     matter halos whereas the low-mass end includes galaxies formed at
     later epochs and active in their ``stellar'' mass formation.
     Finally, the influence of two other physical processes,
     photoionization and non-equipartition processes between
     electrons, ions and neutrals of the cosmological plasma is
     discussed and the modifications on galaxy formation are examined.

      }

   \titlerunning{Thermodynamic evolution of cosmological baryonic gas}
   \authorrunning{Jean-Michel Alimi \& St\'ephanie Courty}
   \maketitle
%

\section{Introduction}

Processes like gravitation, shock heating, radiative cooling,
photoionization and non-equipartition, among others, play a crucial
role in the evolution of the thermodynamic properties of baryonic
matter. As galaxies originate in cold and dense gas regions, any
change in the gas thermodynamic properties should have an effect on
their formation.  This connection is the issue here addressed.
Numerical simulations have the significant advantage of being able to
include a large set of physical processes involved in galaxy
formation. Even if a phenomenological description of this process
needs to be adopted, global galaxy properties are now extensively
studied, like the cosmic star formation rate density, the galaxy
mass/luminosity function, the clustering properties, etc. Such results
use semi-analytical approaches \citep{Somerville99, Cole2000}; hybrid
approaches, combination of semi-analytical and N-body methods
\citep{Kauffmann99, Hatton2003}; and hydrodynamical N-body simulations
using Smooth-Hydrodynamic-Particle \citep{Pearce2001, Weinberg2002},
Lagrangian \citep{Gnedin96} or Eulerian \citep{Cen92} methods. These
complementary computations are all based, but now in a sophisticated
way, on the fundamental ideas that galaxy formation results from the
gas accretion and its cooling \citep{Sunyaev1972, Rees77, Silk77,
  White78} into a network of overdense structures created by the
gravitating dark matter on the large scales of the universe. Moreover,
quantities computed from numerical simulations can now be compared
with the huge amount of available observational data. The galaxy
luminosity function \citep{Binggeli} is widely estimated, in different
surveys, towards fainter magnitude, in several wavelength bands, and
for different classes of galaxies \citep{Madgwick2002}.  Nevertheless
some discrepancies remain: Fig. 1 in \cite{Cross} shows a dispersion
of a factor of 2 at the characteristic luminosity, $L_*$, and a factor
of 10 at $0.01L_*$.  A number of studies attempt to retrieve the
galaxy mass function from the galaxy luminosity function by using
stellar population synthesis models. Numerical simulations take the
opposite approach: their first output is mass and by using stellar
population synthesis models they can provide luminosity. Here we
focus, among other things, on the cosmological evolution of the galaxy
mass function and examine in detail its shape at $z=0$ and $z=1$. We
then derive mass-to-luminosity ratios to compare with the observed
galaxy luminosity function.

As this paper is the second in a series of three, focusing on the
influence of thermodynamics on galaxy formation, the purpose is not to
compute sophisticated models of galaxy formation but rather emphasize
the description of cosmological gas. Hence to keep the model free of
parameters as much as possible, we only consider in the simulations
the dominant physical processes: gravitation, shock heating, radiative
cooling, but neglect e.g. feedback processes. A model of galaxy
formation is also introduced. The first part of the paper examines
global galaxy properties, the cosmic star formation rate density, the
galaxy mass function and the epoch of formation. The main results are
the following: 1. Galaxy formation is a hierarchical process mainly
driven by the amount of available cold gas in the inter-galactic
medium; 2.  The majority of the high-mass galaxies form at early
epochs; 3. The galaxy population at any given redshift is dominated by
a significant fraction of low-mass galaxies formed at early as well as
late epochs.  The halo dark matter mass function is also explored and
a preliminary study of the galaxy distribution inside halos is
presented. The second part of the paper analyzes separately the
influence of photoionization, from ultraviolet background radiation,
and the influence of non-equipartition processes between ions,
neutrals and electrons of the cosmological plasma.  Non-equipartition
has been scrutinized in \cite{Courty} (Paper I) using two numerical
simulations: the first one taking into account non-equipartition
processes and denoted by $S_{3\mathrm T}$ and the second one, denoted
by $S_{1\mathrm T}$, in which equipartition between species is forced.
The former simulation allows each species to carry its own internal
energy whereas the latter one assumes that ions, neutrals and
electrons have the same temperature. That paper concludes that a
significant fraction of the inter-galactic medium (the plasma inside
gravitationally bound structures), that is accreted in not too dense
structures and at temperatures in the range $10^4$--$10^6$ K, is out
of equilibrium and warmer in the $S_{3\mathrm T}$ than in the
$S_{1\mathrm T}$ simulation.  Non-equipartition processes are likely
to be dominant before the end of the reionization epoch. As galaxies
are accreting their gas from the inter-galactic medium in the
temperature range $10^4$--$10^6$ K, this implies an influence of the
non-equilibrium thermodynamics on the galaxy formation process.
Quantifying this change is one of the purposes of this paper. The
third paper in this series (Courty $\&$ Alimi, in preparation), will
quantify how galaxy clustering properties and
the cosmological bias are modified.  \\

This paper is organized as follows. Numerical simulations and the
galaxy formation model are described in section~\ref{simul}.
Section~\ref{galaxy} presents galaxy properties: the cosmic star
formation rate density, the galaxy-like object mass function, and the
epoch of formation. The dark matter halo mass function and the galaxy
distribution inside the biggest mass halos are discussed.  Fits of the
galaxy-like object and dark matter halo mass functions are given.  The
influence of photoionization and non-equipartition processes are shown
in Section \ref{phot} and Section~\ref{s3t}, respectively.
Conclusions are given in Section~\ref{conclu}.

\section{Numerical simulations}
\label{simul}

The simulations were performed with a 3 dimensional
N-body/hydrodynamical code, coupling a Particle-Mesh method for
computing gravitational forces with a Eulerian method
\citep{Teyssier98}. The simulations include shock heating, radiative
cooling, photoionization processes, non-equipartition processes
between the ions, electrons and neutrals of the cosmological plasma,
and galaxy formation. The features of the simulations, analyzed here,
are the following: the $G0$ and $G1$ simulations include shock
heating, radiative cooling and galaxy formation. They only differ in
the resolution. In addition to these processes, the $GP$ simulation
includes photoionization processes and $GNE$ includes
non-equipartition processes but not photoionization. We refer to paper
I for details about shock heating treatment, non-equipartition
processes and the radiative cooling terms. These latter terms include
collisional excitation, collisional ionization, recombination
\footnote{The helium recombination rate of these simulations has an
  incorrect temperature dependence, although the expression for
  $\alpha_{H_e^{++}}$ in Table 1 of Paper I is commonly used in the
  literature. The correct expression should involve the nuclear charge
  of the helium atoms (see \cite{Spitzer78}): $\alpha_{H_e^{++}} =
  3.36\times10^{-10} (1+(T_{e6}/4)^{0.7})^{-1} T_e^{-1/2}
  T_{e3}^{-0.2}$. Since collisional excitation is the dominant net
  cooling term this mistake should have a limited effect on the
  results.}, bremsstrahlung and Compton scattering. To use the same
notations as in paper I, $G0$, $G1$ and $GP$ are $S_{1\mathrm T}$
simulations, in which equipartition between species (ions, neutrals
and electrons) is forced and the cosmological plasma has a single
temperature. The $GNE$ simulation is a $S_{3\mathrm T}$ simulation
with each species having its own internal energy.  \\

The $GP$ simulation takes into account the ionization and heat input
from an ultraviolet background radiation to reproduce conditions after
the reionization epoch. The photoionization and heating rates are
computed from the evolution of the hydrogen and helium densities and
from the spectrum of the ultraviolet background radiation $J(\nu)$.
The radiation is considered a spatially uniform field over the
computational volume. The density evolution equations, Eq. (5) to (7)
in Paper I, are now:
\begin{eqnarray*}
-\beta_{\mathrm H^0} n_\mathrm e n_{\mathrm H^0} + \alpha_{\mathrm H^+} n_\mathrm e n_{\mathrm H^+} - \Gamma_{\mathrm H^0} n_{\mathrm H^0} &=& \frac{\partial n_{\mathrm H^0}}{\partial t}\\ 
-\beta_{\mathrm{He}^0} n_\mathrm e n_{\mathrm{He}^0} + \alpha_{\mathrm{He}^+} n_e n_{\mathrm{He}^+} -\Gamma_{\mathrm{He}^0} n_{\mathrm{He}^0} &=& \frac{\partial n_{\mathrm{He}^0}}{\partial t}\\
\beta_{\mathrm{He}^+} n_\mathrm e n_{\mathrm{He}^+} - \alpha_{\mathrm{He}^{++}} n_\mathrm e n_{\mathrm{He}^{++}} + \Gamma_{\mathrm{He}^+} n_{\mathrm{He}^+} &=& \frac{\partial n_{\mathrm{He}^{++}}}{\partial t}
\end{eqnarray*}
where $n_{\mathrm H^0}$, $n_{\mathrm H^+}$, $n_{\mathrm{He}^0}$,
$n_{\mathrm{He}^+}$, $n_{\mathrm{He}^{++}}$ and $n_e$ are the six
densities of the primordial composition hydrogen-helium cosmological
plasma, and $\beta_i$ and $\alpha_i$ are the ionization and
recombination rates (given in Table 1 in Paper I).  Photo-ionization
rates $\Gamma_i$ are expressed by:
\begin{equation}
\label{Gphot}
\Gamma_i = \int_{\nu_i}^{\infty} \frac{4\pi J(\nu)}{h\nu} \sigma_i(\nu) d\nu\\
\end{equation}
with $i$ denoting the species $\mathrm H^0$, $\mathrm{He}^0$,
$\mathrm{He}^+$, $\sigma_i$ is the effective cross-section for species
$i$ (taken from \cite{Osterbrock}), $h\nu_i$ is the ionization energy for
the species $i$, and $J(\nu)$ is the background radiation intensity.
The photoionization processes are also a heating source (see Eq. (4)
in Paper I) and the heating rates are expressed by:
\begin{equation}
\label{Hphot}
\mathcal{\mathrm H}_i = n_i \int_{\nu_i}^{\infty} \frac{4\pi J(\nu)}{h\nu}
\sigma_i(\nu) (h\nu-h\nu_i) d\nu
\end{equation}

\begin{figure}
\begin{center}
  \includegraphics[height=5.5cm]{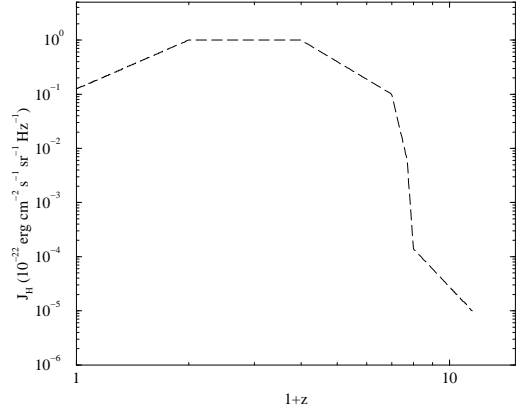}
\caption{Evolution with redshift of the ultraviolet background radiation intensity.}
\label{specphot}
\end{center}
\end{figure}

The shape of the background intensity spectrum is defined by a
function $F(z)$ characterizing the evolution with redshift of the
ultraviolet background radiation \citep{Katz96}:
\begin{equation}
\label{fondUV}
J(\nu) = F(z) \left( \frac{\nu}{\nu_\mathrm H} \right)^{-1} \ \dt{erg} \ \dt{cm}^{-2} \ \dt{s}^{-1} \ \dt{sr}^{-1} \ \dt{Hz}^{-1}
\end{equation}
where $h \nu_\mathrm H=13.598 \ \dt{eV}$ is the hydrogen ionization
threshold.  Since this evolution is little-known at high redshift, we
estimate $J(\nu)$ from observational measurements \citep{Scott} and
from numerical works \citep{Gnedin2000}. This includes the decline of
the ultraviolet background radiation intensity observed between $z
\sim 1$ and $z=0$, the sharp increase before $z \sim 6$ and the
shallow evolution before $z \sim 7$. We start the reionization at
redshift $10.5$ with the bulk of the transition between $z=7$ and
$z=6$. The function $F(z)$ is plotted in Fig.~\ref{specphot} and its
expression, in units of $J_0= 10^{-22}\ \dt{erg} \ \dt{cm}^{-2}
\dt{s}^{-1} \ \dt{sr}^{-1} \ \dt{Hz}^{-1}$, is:
\begin{equation}
F(z) = 
\begin{cases}
10^{(A1log(1+z)+B1)} & 11.5 \ge 1+z > 8 \\
10^{(A2log(1+z)+B2)} & 8 \ge 1+z > 7.7 \\
10^{(A3log(1+z)+B3)} & 7.7 \ge 1+z > 7 \\
J_0(4/1+z)^{4.11} & 7 \ge 1+z > 4 \\ 
J_0 & 4 \ge 1+z > 2 \\ 
J_0(2/(1+z))^{-3} & 2 \ge 1+z \\ 
\end{cases}
\end{equation}
with $A1 = -7.27, A2 = -98.32, A3 = -29.52, B1 = -19.28, B2 = 62.94,
B3 = 1.94$. 
\\

\begin{figure}[h]
\begin{center}
  \includegraphics[height=5.8cm]{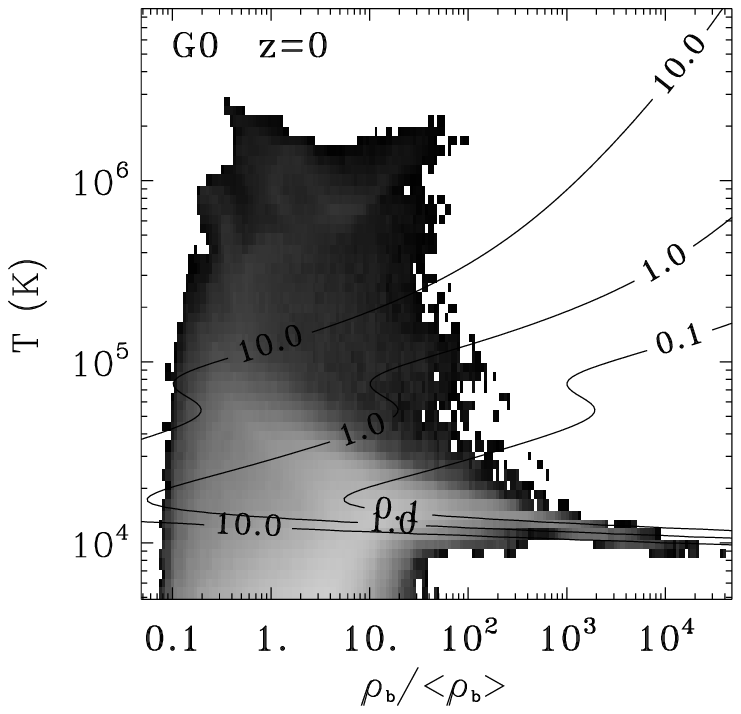}
  \includegraphics[height=5.8cm]{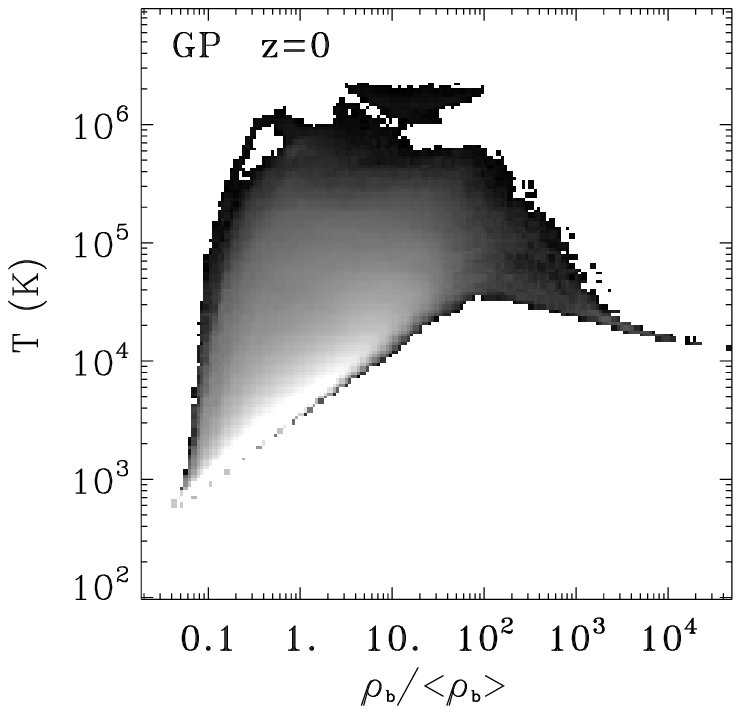}
  \includegraphics[height=5.8cm]{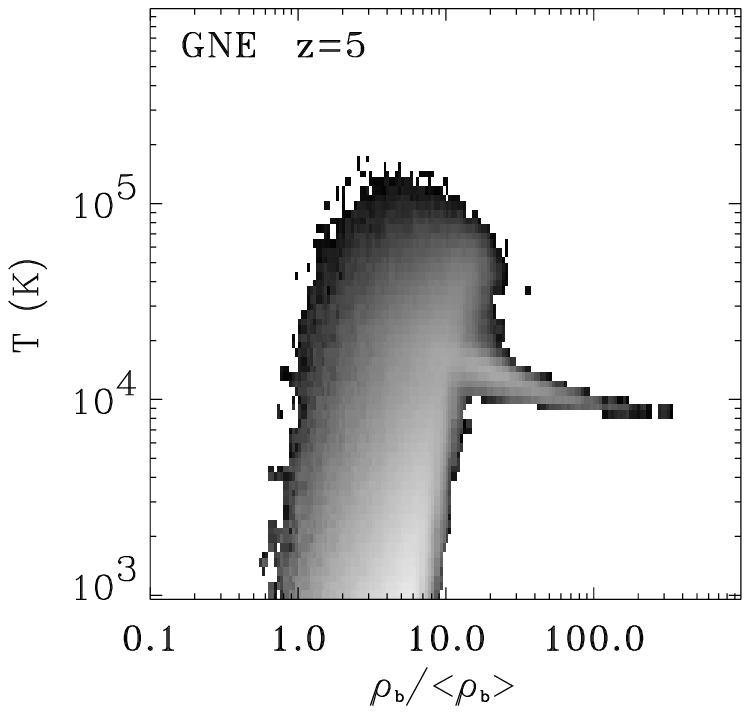}
\caption{Isocontours of the baryonic mass fraction per interval of the
  baryonic density contrast and per interval of temperature computed
  for the $G1$ and $GP$ simulations at $z=0$ and for $GNE$ at $z=5$
  (top, middle and bottom panels, respectively). The increase in the
  mass fraction scales from dark to light. Note that $G1$ and $GNE$
  are, for these plots, computed without galaxy formation. Solid
  curves on the top panel show the ratios $t_{cool}/t_{ff}=0.1$, 1,
  10.}
\label{iso}
\end{center}
\end{figure}

As it is useful for the discussion in this paper, we now compare the
gas distribution in the three kinds of simulations. Figure \ref{iso}
illustrates the baryonic mass fraction in temperature-density
diagrams; the top and bottom panels are extracted from Paper I. The
top panel is computed with a simulation including hydrodynamical
shocks and radiative cooling. The middle panel includes
photoionization processes and shows that they are dominant over
cooling processes only in low and middle dense regions (see also
\cite{Weinberg97}): low density regions are heated up to temperatures
between $10^3$ and less than $10^4$ K. The competition between
photoionization heating and cooling due to adiabatic expansion results
in the concentration of the gas on a slope $T =T_0
(\rho/\bar{\rho})^\gamma$ with $T_0 \sim 6.10^3$ K and $\gamma$ around
0.6 (\cite{Katz96, Gnedin1997}). The bottom panel displays the gas
distribution in the $GNE$ simulation and shows a non-negligible warm
gas fraction. As the influence of non-equipartition processes is
larger at high redshift, the isocontours are plotted at $z=5$ (see
discussion in Paper I).

Although differences exist in the low-density regions, altogether the
three gas distribution diagrams show a common feature: a peak in the
high density, cold region. This gas being the reservoir for galaxy
formation, we thus present in the first part of this paper galaxy
properties in a simulation only including the dominant processes
involved in galaxy formation, namely hydrodynamical shocks and
radiative cooling. The influence of photoionization and
non-equipartition processes will be considered in separate sections.
\\

We now turn to the description of the galaxy formation model.
Numerical simulations of large scale structure formation currently do
not allow for the formation of objects beyond the scale of a grid
cell, a few times $10^6 \ \mathrm M_{\odot}$, this mass being much
larger than the mass of a single star. The problem is bypassed by
considering the physical conditions needed to form a galaxy. The most
important condition is that the gas cloud is collapsing, meaning that
the cooling time is less than the dynamical time or the free fall time
\citep{Rees77}. The top panel of Fig.~\ref{iso} illustrates in a
temperature-baryonic density diagram that dense and cold gas regions
are located inside the iso-contour $t_\mathrm{cool}/t_\mathrm{ff}=1$.
The galaxy formation model then consists of the identification of the
gas satisfying this criteria. A fraction of the baryonic matter is
turned into a ``stellar'' particle describing the amount of stellar
mass produced during the process of galaxy formation. Galaxy-like
objects are then defined by a collection of this ``stellar'' mass (the
term ``galaxy-like object'' has already been used in \cite{Evrard},
although describing high-density contrast baryonic clumps). To make
sure that gas regions giving birth to galaxies are correctly
identified we add other criteria described below. The purpose of this
paper being to show how modifications of the gas thermodynamics have
an effect on the galaxy formation process, we then deliberately keep
the number of free parameters low. This model, although very simple,
gives consistent results between the properties of the galaxy-like
objects in the simulation and their observational counterparts.

To express the condition $t_\mathrm{cool} < t_\mathrm{ff}$, we define
the cooling timescale $t_\mathrm{cool}$, computed from the internal
energy variation of the gas $E/ \dot E$, and the dynamical time or
free fall time:
\begin{equation}
t_\mathrm{ff} = \sqrt\frac{3 \pi}{32 G \rho}
\end{equation}
One of the other conditions requires that the size of the gas cloud must be
less than the Jean's length given by:
\begin{equation}
\lambda_\mathrm J = c_\mathrm s \left( \frac{\pi}{G \rho} \right)^{1/2}
\end{equation}
Note that the total matter density, including dark matter, baryonic
matter and "stellar" particles, is used in the expressions of the
dynamical time and the Jean's length. It is clear that the Jeans
criterion is reliable only down to the mass resolution, since the size
of gas clouds itself is limited by the spatial resolution of the
simulation. A third condition is that the gas must be in a converging
flow: $\nabla \cdot \vec v < 0$. Finally the baryonic density
contrast, $\delta_\mathrm B \equiv (\delta \rho /\bar{\rho})_\mathrm
B$, must be higher than a threshold $(1+\delta_B)_s$.  This is taken
to be the value of the baryonic density contrast at the turnaround,
5.5, computed in the top-hat collapse spherical model
\citep{Padmanabhan}.

To estimate the amount of ``stellar'' mass formed, we express the
variation of the baryonic mass as the ratio between the available
baryonic mass $m_\mathrm B$ and a characteristic timescale $t_*$:
\begin{equation}
-\frac{\textrm{d}m_\mathrm B}{\textrm{d}t}=\frac{m_\mathrm B}{t_*}
\end{equation}
The integration of this expression on a timestep $\Delta t = t-t_0$
gives:
\begin{equation}
m_\mathrm B(t) = m_\mathrm B(t_0) \textrm{exp}(-\frac{\Delta t}{t_*})
\end{equation}
where $m_\mathrm B(t_0)$ is the baryonic mass initially present. The
``stellar'' mass formed is then:
\begin{equation}
\label{mstar}
m_* = m_\mathrm B(t_0) - m_\mathrm B(t) \simeq m_\mathrm B(t_0) \frac{\Delta t}{t_*}
\end{equation}

Then in each cell, checking the four criteria described above, a
fraction of the gas is turned into a ``stellar'' particle. Each of
these particles carries its mass $m_*$ and its epoch of formation
given by the scale factor $a_*$. The mass $m_*$ is computed using
Eq.~(\ref{mstar}) with $m_\mathrm B(t_0)$ the baryonic mass enclosed
within the grid cell at each timestep and with the characteristic time
$t_*=\dt{max}(t_\mathrm {ff},10^8 \ \dt{yr})$. Figure~\ref{fctmGL} in
Appendix \ref{A1} shows the evolution in redshift of the ``stellar''
particle mass function for the different simulations: the ``stellar''
particle mass ranges between a few times $10^5$ and $\sim 2.10^8$
$\mathrm M_{\odot}$. The ``stellar'' particles are involved in the
computation of the gravitational potential and their evolution is
treated in the same way as the collisionless dark matter.

At any redshift two catalogs of objects are created: one consisting of
dark matter halos and one of galaxy-like objects. Halos and galaxies
are defined by grouping either dark matter particles or ``stellar''
particles with a Friend-of-Friend algorithm.  This algorithm joins
together all particles separated by a distance proportional to the
link parameter $\eta$. We take $\eta=0.2$. We exclude from the dark
matter halo catalog groups with less than 10 particles. This threshold
is denoted $M_\mathrm{min}$. But the galaxy-like object catalog is
allowed to include objects with a lower $M_\mathrm{min}$, meaning that
each ``stellar'' particle is considered a galaxy-like object. The
influence of these two parameters on the galaxy mass function, $\eta$
and $M_\mathrm{min}$,
are discussed in Appendix~\ref{sensitivity}.  \\

Unless otherwise stated, the results of this paper are given for a
$\Lambda-$cold dark matter model ($\Lambda-CDM$). The parameters of
the simulations are: $H_0=70.\ \dt{km} \ \dt{s}^{-1} \ \dt{Mpc}^{-1}$,
$\Omega_\mathrm K=0.$, $\Omega_\mathrm m=0.3$, $\Omega_{\Lambda}=0.7$,
$\Omega_\mathrm b=0.02h^{-2}$ with $h=H_0/100$. The initial density
fluctuation spectrum uses the transfer functions taken from
\cite{Bardeen} with a shape parameter given by \cite{Sugiyama}.  The
fluctuation spectrum is normalized to COBE data \citep{Bunn} leading
to a filtered dispersion at $R=8 \ h^{-1} \ \textrm{Mpc}$ of
$\sigma_8=0.91$. The number of dark matter particles is $N_\mathrm
p=256^3$ and the number of grid cells is $N_\mathrm g=256^3$. Three
computational volumes are used, described, as well as the simulation
parameters, in Table~\ref{param}.

\begin{table}
\begin{center}
\begin{tabular}{lcccccc}
\\
Simulation & $L_\mathrm{box}$ & $dx$ & $M_\mathrm{dm}$ & $M_\mathrm{bm}$  \\
& ($h^{-1}$Mpc) & ($h^{-1}$kpc) & ($\mathrm M_{\odot}$) & ($\mathrm M_{\odot}$) \\
\hline
\\
$G0$ & 32. & 125. & $2.01\times 10^8$ & $3.09\times 10^7$  \\
$G1$, $GNE$ & 16. & 62.5 & $2.51\times 10^7$ & $3.87\times 10^6$  \\
$GP$ & 11. & 43. & $8.17\times 10^6$ & $1.25\times 10^6$  \\
\end{tabular}
\caption{\label{param} Parameters of the simulations. $L_\mathrm{box}$
  is the comoving length of the computational volume, $dx$ is the
  spatial resolution of the grid, $M_\mathrm{dm}$ is the mass of the dark
  matter particle and $M_\mathrm{bm}$ is the initial baryonic mass enclosed
  within a grid cell.}
\end{center}
\end{table}

\begin{figure}
\begin{center}
\includegraphics[height=8cm, angle=-90]{fig3.ps}
\caption{Evolution with redshift of the cosmic ``stellar'' mass formation
  rate density for the $G1$ (dashed line), $G0$ (dot-dashed line) and
  $GP$ (dotted line) simulations. The observational data of the star
  formation rate density are overplotted for our cosmology: H$\alpha$
  data: \cite{Gallego} (filled circle), \cite{Tresse} (filled
  triangle), \cite{Yan} (filled star), \cite{Hopkins} (hollow
  diamond); UV data: \cite{Connolly} (open square), \cite{Treyer}
  (open triangle), \cite{Steidelb} (open circle), \cite{Sullivan}
  (diamond); FIR data: \cite{Rowan} (heavy cross), \cite{Flores}
  (cross); 1.4 GHz data: \cite{Condon} (filled square), \cite{Haarsma}
  (six-pointed star), \cite{Serjeant} (open star). UV data and data by
  \cite{Gallego} are corrected for extinction.}
\label{sfr}
\end{center}
\end{figure}

\begin{figure}
\begin{center}
\includegraphics[height=8cm, angle=-90]{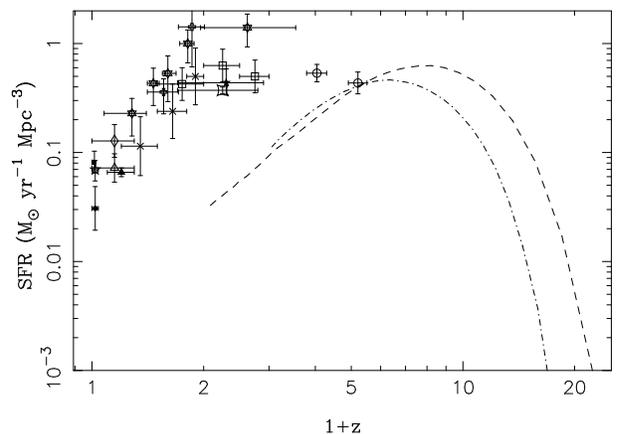}
\caption{Same as Fig.~\ref{sfr} but for a $SCDM$ scenario, for the $G1$ (dashed line) and $G0$ (dot-dashed line) simulations. The observation points are the same as described in Fig.~\ref{sfr} but overplotted for a $SCDM$ cosmology.}
\label{sfr_scdm}
\end{center}
\end{figure}

\section{Galaxy properties}
\label{galaxy}

\subsection{Cosmic ``stellar'' mass formation rate density}

Figure~\ref{sfr} shows the redshift evolution of the ``star''
formation rate (SFR) density for the $G0$ and $G1$ simulations. We
compute the amount of ``stellar'' mass formed per year and per unit of
volume. The SFR density shows a strong decrease at low redshift and
peaks around redshift $\sim 3$ in $G0$ and 2.5 in $G1$. However, the
ratio between the SFR at the peak and the SFR at $z=6$ is lower than
between the SFR at the peak and at $z=0$,
$SFR(z=z_{\mathrm{peak}})/SFR(z=6) \sim$ 3.2 against
$SFR(z=z_{\mathrm{peak}})/SFR(z=0) \sim$ 10 in the $G0$ simulation
(these values become 2.8 and 6.3 in $G1$, respectively).  This trend
is related to the hierarchical nature of cold dark matter models. At
high redshift baryonic matter is accreted in dark matter potential
wells and cools to form galaxy material. Then low-mass structures
merge together to form larger mass units. Additional gas accreted is
then shock heated towards higher temperatures leaving insufficient
time for the gas to cool and condense. This results in the decrease in
the star formation rate at low $z$.

Another illustration of the influence of the large scale structures on
galaxy formation comes from the comparison with a different
cosmological scenario. We run the same simulations but now computed
for a standard cold dark matter model ($SCDM$) with the following
parameters: $H_0=50.\ \dt{km} \ \dt{s}^{-1} \ \dt{Mpc}^{-1}$,
$\Omega_\mathrm K=0.$, $\Omega_\mathrm m=1.$, $\Omega_{\Lambda}=0.$,
$\Omega_\mathrm b=0.02h^{-2}$ with $h=H_0/100$. As in the
$\Lambda-CDM$ scenario, the fluctuation spectrum is normalized to COBE
data, giving $\sigma_8=1.1$. For computational reasons the $SCDM$ $G1$
simulation is only performed up to $z=1$ and the $G0$ up to $z=2$. The
density fluctuation power spectrum has more power on small scales than
the $\Lambda-CDM$ model, resulting in a larger amount of cold gas at
high redshift. Differences between these two models are strikingly
illustrated by the cosmic star formation rate density
(Fig.~\ref{sfr_scdm}). The higher amount of dark matter in the $SCDM$,
resulting in a slightly larger $\sigma_8$ than in the $\Lambda-CDM$
scenario, creates deep potential wells at high redshift and as a
result the peak of the SFR density is reached at $z=6$, or even higher
$z$, leading to a steep decrease at low redshift.

\begin{figure}
\begin{center}
\includegraphics[height=7.5cm, angle=-90]{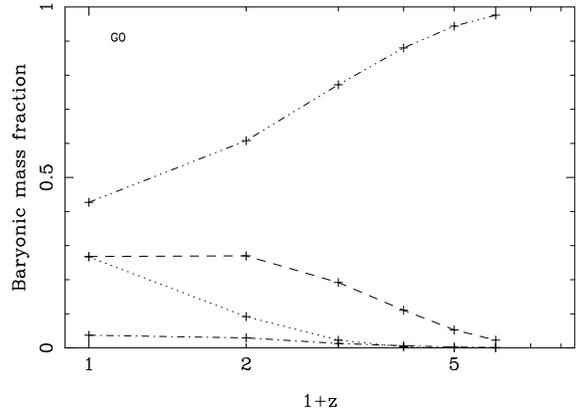}
\caption{Evolution with redshift of the baryonic mass
  fractions computed in different temperature ranges, for the $G0$
  simulation but without galaxy formation: $T<9.10^3$ K (``diffuse'',
  dot-dot-dashed line), $9.10^3 \leq T< 2.10^4$ K (``cold'', dashed
  line), $2.10^4 \leq T< 5.10^5$ K (``warm'', dot-dashed line), $T \geq
  5.10^5$ K (``hot'', dotted line).}
\label{fracL32}
\end{center}
\end{figure}

\begin{figure}
\begin{center}
\includegraphics[height=7.5cm, angle=-90]{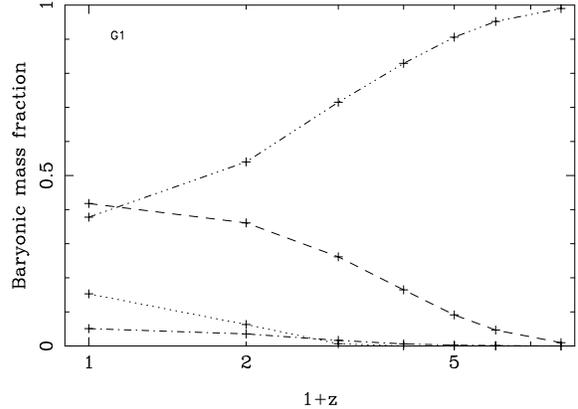}
\caption{Same as Fig.~\ref{fracL32}, but for the $G1$ simulation without galaxy formation.}
\label{fracL16}
\end{center}
\end{figure}

Figure~\ref{sfr} shows that the SFR density depends on the spatial
resolution of the simulations: the highest resolution simulation,
$G1$, has a SFR density amplitude higher at low redshift than the $G0$
simulation. Decreasing the box length allows to include in the
computational volume density fluctuations with lower wavelengths,
resulting in a higher mass fraction of available cold gas (similar
trends are discussed in \cite{Weinberg99} and \cite{Ascasibar}).
Figures~\ref{fracL32} and~\ref{fracL16} display, for the same
simulations as $G0$ and $G1$ but without galaxy formation, the
evolution with redshift of the baryonic mass fraction in different
ranges of temperature, corresponding to the main phases of the
inter-galactic medium: the ``diffuse'' phase with $T<9.10^3$ K, the
``cold'' phase with temperature in the range $9.10^3$--$2.10^4$ K, the
``warm'' phase in the range $2.10^4$--$5.10^5$ K and the ``hot'' phase
at $T \geq 5.10^5$ K. The increase in the amount of cold gas with
decreasing redshift is less dramatic for the $G0$ simulation: between
$z=5$ and $z=0$, the cold gas mass fraction goes from 5.2$\%$ to
26$\%$ whereas it goes from 4.6$\%$ to 41$\%$ in $G1$. However
changing the resolution does not change the evolution of the SFR
density and we are less interested in making quantitative estimates
than analyzing qualitatively the process of galaxy formation.

Observational data of the star formation rate density are overplotted
in Fig.~\ref{sfr}. These data show a great disparity and the amplitude
of the SFR density is expected to change due to different
observational bias corrections \citep{Hopkins2001}. Nonetheless a
strong decrease is observed at low redshift. Moreover observations do
not currently agree whether the SFR density peaks after $z=1$ or
reaches a plateau at higher redshift. Both star formation rate
densities in our simulations reproduce the slope at low redshift and
the amplitude obtained in the $G1$ simulation is consistent with
observations (contrary to the $SCDM$ scenario showing a decrease at
earlier epochs). This agreement and the previous discussion point out
that galaxy formation is mainly driven by the amount of available cold
gas enclosed in gravitationally bound structures, the strongest
constraint of the galaxy formation model lying in the condition
$t_\mathrm{cool} < t_\mathrm{ff}$ (see also \cite{Pearce2001}).

\subsection{Galaxy-like object mass function}

The rest of Section~\ref{galaxy} is now devoted to the description of
our two galaxy catalogs and examines an important property of galaxy
formation, the galaxy-like object mass function. Discussions about
their epoch of formation and their location inside dark matter halos
will follow.

\begin{figure}
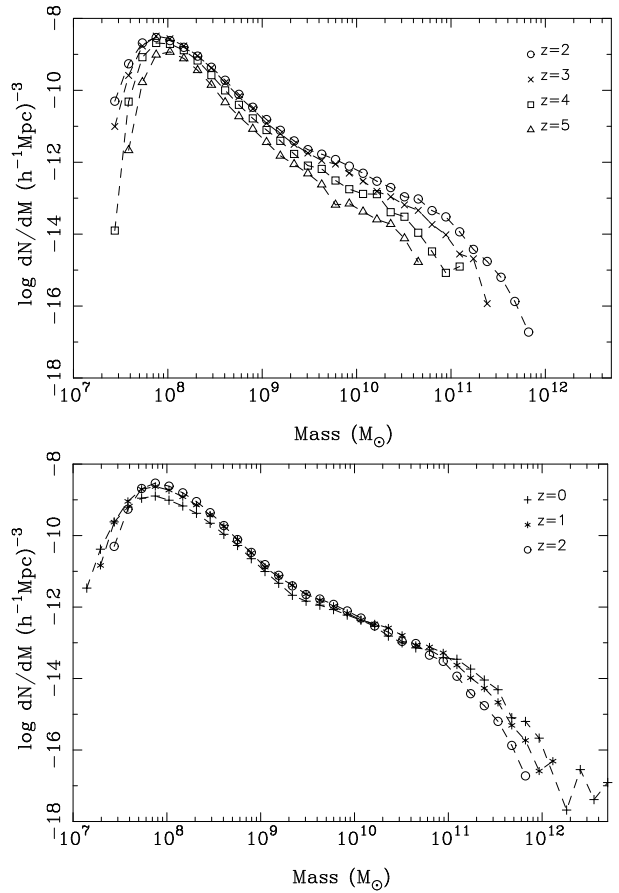

\begin{center}
\includegraphics[height=8cm, angle=-90]{fig5a.ps}
\includegraphics[height=8cm, angle=-90]{fig5b.ps}
\caption{Galaxy-like object mass function at different redshifts for the $G0$ simulation.}
\label{fctmGG_L32}
\end{center}
\end{figure}

Figure~\ref{fctmGG_L32} shows the cosmological evolution of the mass
function of the galaxy-like objects in the $G0$ simulation. The
catalog covers a wide range a mass, from $10^8 \ \mathrm M_{\odot}$ up
to $10^{12} \ \mathrm M_{\odot}$. As we do not know if the decrease in
the range $M < 10^8 \ \mathrm M_{\odot}$ is a numerical effect (see
also \cite{Murali}), we discuss in the following the shape of the mass
function for $M > 10^8 \ \mathrm M_{\odot}$. At any redshift the
comoving number density of objects per bin of mass increases as the
mass decreases. This trend is different from the stellar particle mass
functions (Fig.~\ref{fctmGL}). The galaxy mass function shows a clear
evolution from $z=5$ to $z=0$: more and more objects of high-mass are
created, increasing the mass range of galaxies towards the high-mass
end. Since galaxy formation results from a hierarchical process, the
mass of bigger objects increases with decreasing redshift, shifting
the knee of the mass function towards larger mass. On the other hand
the low-mass end of the mass function shows an increase as redshift
decreases to only $z=2$. At lower redshift the trend is inverted with
a decrease in the number density per bin of mass of low-mass objects.

\begin{figure*}
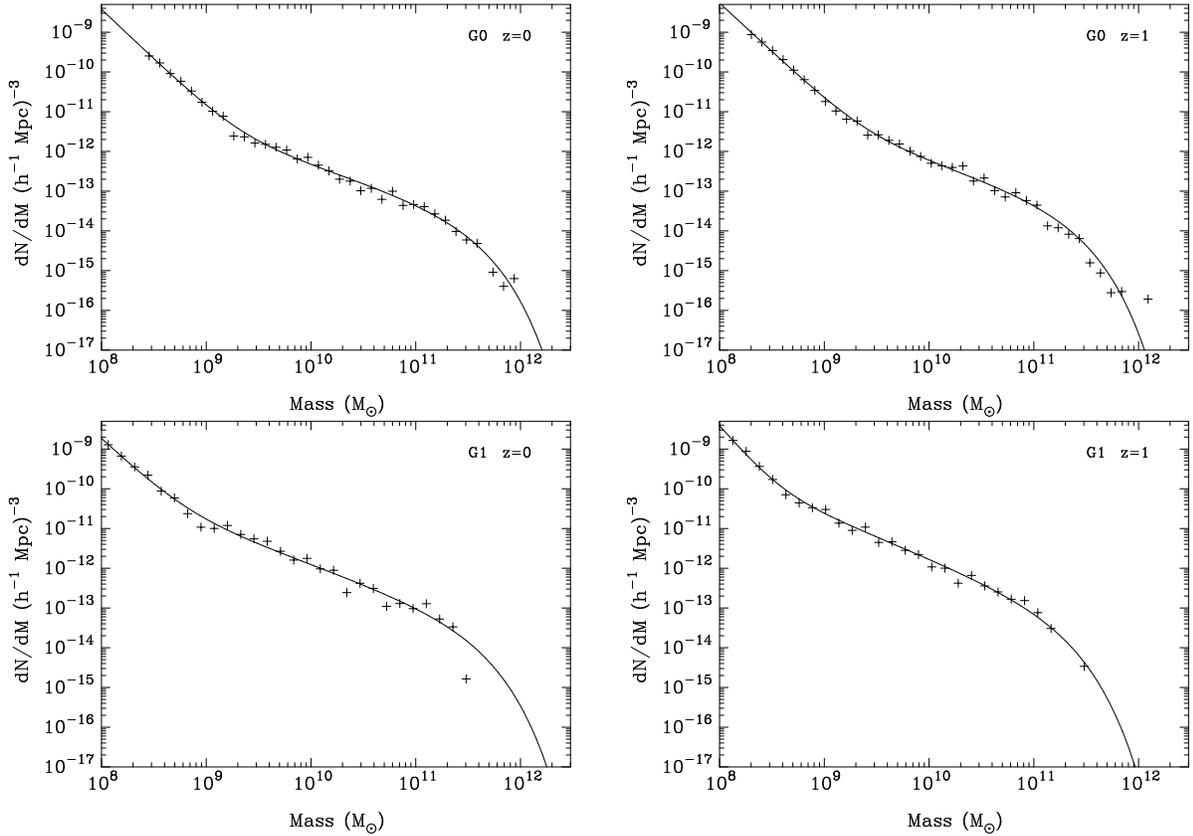

\begin{center}
\includegraphics[height=7.5cm, angle=-90]{fig6a.ps}
\hspace{.5cm}
\includegraphics[height=7.5cm, angle=-90]{fig6b.ps}
\vspace{.5cm}
\includegraphics[height=7.5cm, angle=-90]{fig6c.ps}
\hspace{.5cm}
\includegraphics[height=7.5cm, angle=-90]{fig6d.ps}
\caption{Analytical fits (solid curve) of the galaxy-like object mass
  function (cross) for the $G0$ and $G1$ simulations at $z=0$ and
  $z=1$. Fits are computed using Eq.~(\ref{fit}).}
\label{fctmGG_fit}
\end{center}
\end{figure*}

On the whole the galaxy mass function shows a characteristic shape: a
strong decrease at the high-mass end beyond a characteristic mass,
preceded by a shallower slope in the intermediate mass range, between
$10^9$--$10^{11} \ \mathrm M_{\odot}$, and a steeper slope at the
low-mass end up to $10^8 \ \mathrm M_{\odot}$. Note the sharp
transition at $z=0$ around a few times $10^9 \ \mathrm M_{\odot}$,
shown more clearly in Fig.~\ref{fctmGG_fit}.
\\

Figure~\ref{fctmGG_L32} presents one of the main results of the paper:
the galaxy mass function is significantly dominated by a low-mass
galaxy population, $M < 10^{10} \ \mathrm M_{\odot}$, whatever the
redshift is. This population can be linked to the observed faint
luminosity galaxy population (\cite{Loveday, Norberga}, and references
therein).  In fact this faint population covers a wide variety of
galaxies: galaxies with or without emission lines \citep{Zucca},
galaxies characterized by a significant star formation activity
\citep{Lin, Loveday99}, dwarf galaxies of morphological and spectral
late types \citep{Marzke94, Marzke98}, low surface brightness galaxies
\citep{Sprayberry}, blue compact objects \citep{Guzman1997}. The
discussion below about the epoch of formation of the galaxy-like
objects will show that, in the simulations also, the low-mass end of
the galaxy mass function includes an inhomogeneous population. This
population of faint galaxies is important for galaxy evolution, since
by their number they are likely to contribute to the cosmic star
formation rate density. Their relationship to the environment and
their implication in galaxy mergers still needs to be addressed.

\begin{table}
\begin{center}
\begin{tabular}{l l l l}
Reference & $\phi^*$ & $\alpha$ & $\mathcal{M}_*$ \\
\hline
\\
$B$ band & \\
\cite{Loveday92} & 0.014 & $-0.97$ & $-20.27$ \\
\cite{Marzke94} & 0.04 & $-1.0$ & $-20.02$ \\
\cite{Ellis96} & & \\
\cite{Loveday97} & 0.0154 & $-0.94$ & $-20.42$ \\
& & $-1.82 (\beta)$ & $-14.84 (\mathcal{M}_t)$ \\
\cite{Zucca} & 0.02 & $-1.22$ & $-20.38$ \\
\cite{Zucca} $(b)$ & 0.021 & $-1.16$ & $-20.34$ \\
& & $-1.57 (\beta)$ & $-17.76 (\mathcal{M}_c)$ \\
\cite{Ratcliffe} & 0.017 & $-1.04$ & $-20.45$ \\
\cite{Norberga} & 0.0168 & $-1.21$ & $-20.43$ \\
\\
$K$ band & \\
\cite{Glazebrook} & 0.029 & $-1.04$ & $-23.5$ \\
\cite{Gardner97} & 0.0166 & $-0.91$ & $-23.9$ \\
\cite{Loveday2000} & 0.012 & $-1.16$ & $-24.35$ \\
\cite{Kochanek} & 0.0116 & $-1.09$ & $-24.16$ \\
\cite{Cole2001} & 0.0116 & $-0.93$ & $-24.13$ \\
\end{tabular}
\end{center}
\caption{\label{litt} Fit parameters of the galaxy luminosity function extracted from the literature for different surveys using a standard Schechter function (Eq.~(\ref{schechter})). Magnitudes are computed for $h=0.7$. The normalization parameter is expressed in $h^3\mathrm{Mpc}^{-3}$. A two power-law Schechter function is used in \cite{Loveday97} with a correction for the $\beta$ parameter in \cite{Loveday}. \cite{Zucca} describes the faint part of the galaxy luminosity function by a power-law introducing the $\beta$ slope and the magnitude $\mathcal{M}_\mathrm c$ (line noted $(b)$).}
\end{table}

The galaxy luminosity function is generally fitted by a
standard Schechter function combining a power-law and an
exponential function at the bright end \citep{Schechter}:
\begin{equation}
\label{schechter}
\phi(L) = \frac{\phi_*}{L_*} \left( \frac{L}{L_*} \right)^{\alpha}
\dt{exp} \left( -\frac{L}{L_*} \right)
\end{equation}
where $\alpha$ is the slope of the power-law, $L_*$ is the
characteristic luminosity at the break, and $\phi_*$ is a
normalization parameter. Table~\ref{litt} lists parameter fits,
extracted from the literature, of the galaxy luminosity function for
different surveys. In order to differentiate mass from magnitude, the
symbol $\mathcal{M}$ is used for magnitude in this table and
throughout the paper. As pointed out in the introduction, discrepancies
remain between surveys, enlightening the fact these surveys cover
different galaxy populations. Observational results show that the
galaxy luminosity function is different for galaxy populations
selected by color, morphology or environnement.  The 2$d$F survey, for
instance, shows how the shape of the galaxy luminosity function
changes when galaxies are divided according to their star formation
activity \citep{Folkes1999}.

Most of these surveys use a single power-law
Schechter function and the $\alpha$ slope is generally found to be in
the range $-0.9$ and $-1.2$. A number of estimates of the galaxy
luminosity function conclude that it cannot be described correctly, in
the entire range of luminosity and specially at $\mathcal{M}_{B} >
-15$, by a single power-law Schechter function \citep{Loveday}. Two
works, reported in Table~\ref{litt}, use a non-standard Schechter
function to parameterize the galaxy luminosity function: a two
power-law Schechter function described below, is used in
\cite{Loveday97} and two analytical functions are used in
\cite{Zucca}, a standard Schechter function and a power-law at the
faint luminosity end. Both report slopes in the faint luminosity range
of $-1.82$ and $-1.57$, respectively.

The mass function of the galaxy-like objects in Fig.~\ref{fctmGG_L32}
clearly suggests similar conclusion. Therefore to account for the
low-mass galaxy population we choose a two power-law Schechter
function now combining a standard Schechter function with a $\beta$
power-law at the low-mass end \citep{Loveday97}:
\begin{equation}
\label{fit}
\frac{\ud N}{\ud M} =
\frac{\phi_*}{M_*} \left( \frac{M}{M_*} \right)^{\alpha} \dt{exp}
\left( -\frac{M}{M_*} \right) \left[ 1+
\left(\frac{M}{M_\mathrm t}\right)^{\beta} \right]
\end{equation}
where $\phi(M)=\ud N/\ud M$ is the numerical density of objects per
interval of mass and per unit of volume, $\phi^*$, $M_*$ and $\alpha$
are analogous parameters as used in Eq.~(\ref{schechter}), and
$M_\mathrm t$ is the transitional mass between the two power-laws.
The mass function of the galaxy-like objects is fitted, in the range
$M > 10^8 \ \mathrm M_{\odot}$, using a least square method weighted
by the mass function itself to ensure the statistical reliability of
the result.  Figure \ref{fctmGG_fit} and Table~\ref{fitGG} give fits
and parameters determined at $z=0$ and $z=1$ for the $G0$ and $G1$
simulations. At $z=0$ the characteristic mass at the high-mass end is
roughly similar for the two resolutions, around $2.10^{11} \ \mathrm
M_{\odot}$. The $\alpha$ slope is quite shallow, around or higher than
$-1$, whereas the $\beta$ slope is less than $-1.5$. These values are
consistent with the ones obtained in different surveys
(Table~\ref{litt}). It is quite remarkable that such a simple galaxy
formation model reproduces the shape of the galaxy luminosity
function. The $\alpha$ slope does not show strong evolution between
$z=0$ and $z=1$, contrary to the characteristic mass decreasing as the
redshift increases. The transitional mass and the normalization
parameter show clearly the influence of resolution: the $\alpha$
power-law extends on a higher mass range for the highest resolution,
changing the transitional mass: $M_\mathrm t \sim 6.10^8$ $\mathrm
M_{\odot}$ for the $G1$ simulation whereas $M_\mathrm t \sim 2.10^9$
$\mathrm M_{\odot}$ in $G0$.  The bottom panel in Fig.~\ref{fctm_comp}
shows a comparison between both resolutions at $z=0$. Indeed a higher
resolution allows the formation of lower mass galaxy-like objects. The
mass functions tend to be steeper in the intermediate range at $z=1$,
and the characteristic mass $M_*$ decreases. We will return to these
differences at the end of this section.

\begin{table}
\begin{center}
\begin{tabular}{c c c c c c c c}
 & $\phi_*$ & $M_*$ & $\alpha$ & $M_\mathrm t$ & $\beta$ \\
\hline
\\
$G0$ \\
$z=0$ & 0.00720 & $2.53 \times 10^{11}$ & $-0.86$ & $2.13 \times 10^9$ & $-1.65$ \\
$z=1$ & 0.00824 & $1.67 \times 10^{11}$ & $-0.87$ & $2.55 \times 10^9$ & $-1.59$ \\
\\
$G1$ \\
$z=0$ & 0.0144 & $2.64 \times 10^{11}$ & $-0.96$ & $6.35 \times 10^8$ & $-1.52$ \\
$z=1$ & 0.0146 & $1.29 \times 10^{11}$ & $-1.08$ & $3.69 \times 10^8$ & $-2.02$  \\
\end{tabular}
\end{center}
\caption{\label{fitGG} Fit parameters (Eq.~(\ref{fit})) computed for the galaxy-like object mass function (Fig.~\ref{fctmGG_fit}) for the $G0$ and $G1$
simulations at $z=0$ and $z=1$. The characteristic masses are expressed in $\mathrm M_{\odot}$ and the normalization parameter in $h^3 \mathrm{Mpc}^{-3}$.}
\end{table}

The integration of the galaxy mass function gives the mean mass
density, $j_\mathrm M = \int \phi(M) M \ud M$. Using the fits
displayed in Table~\ref{fitGG} we compute $j_M$ over the range $M>10^8
\ \mathrm M_{\odot}$.  At $z=0$ this quantity is $8.73 \times 10^8$
and $1.84 \times 10^9 \ h \ \mathrm M_{\odot} \ \mathrm{Mpc}^{-3}$ for
the $G0$ and $G1$ simulations, respectively.  At $z=1$ the mean mass
densities are lower: $6.9 \times 10^8$ and $9.9 \times 10^8 \ h \ 
\mathrm M_{\odot} \ \mathrm{Mpc}^{-3}$ for the same simulations,
respectively. Normalized to the critical density parameter at $z=0$,
$\rho_{c,0} =2.67 \times 10^{11} \ h^2 \ \mathrm M_{\odot} \ 
\mathrm{Mpc}^{-3}$, the stellar density parameter decreases from
$\Omega_*=0.00451$ to 0.00357 between $z=0$ and $z=1$ for the $G0$
simulation, and from $\Omega_*=0.00954$ to 0.00511 for $G1$. The
values at $z=0$ can be compared to \cite{Fukugita}, reporting a
central value of $\Omega_*=0.0035$ (see their Table 3) which accounts
for stars in spheroids, disks and irregulars.

In order to compare the characteristic masses $M_*$ and $M_\mathrm t$
with the characteristic magnitudes of the galaxy luminosity function,
we use the mean mass densities to derive the mass-to-light ratios of
the galaxy-like object catalogs.  We express the mass-to-light ratio
as $\gamma=j_\mathrm M/j_\mathrm L$ where $j_\mathrm L$ is the mean
luminosity density, $j_\mathrm L = \int_0^\infty \phi(L) L \ud L$.
Note that this ratio is a ``stellar'' mass-to-light ratio since the
galaxy-like objects enclose only ``stellar'' material.  The mean
luminosity density is taken to be $2.5 \times 10^8$ and $5.10^8$ $h \ 
L_{\odot} \ \mathrm{Mpc}^{-3}$ in the $b_J$ and $K$ bands,
respectively \citep{Cole2001}. Table~\ref{magz0} displays the
mass-to-light ratios in each band and for the two resolutions. Values
are higher in the $B$ band and also for the highest resolution
simulation. They are between 1.7 and 7.4, consistent with
observational results (\cite{Fukugita} give mass-to-light ratios of
6.5 for spheroids and 1.5 for disks). Recall that these mass-to-light
ratios are computed over an extensive range of mass, and that this
quantity is likely to depend on mass.

\begin{table}
\begin{center}
\begin{tabular}{c c c c c}
Simulation & $\gamma$ & $\mathcal{M}_*$ & $\mathcal{M}_\mathrm t$  \\
\hline
\\
$G0$ & 3.5 & $-21.8$ & $-16.6$ & $B$ band \\
$G0$ & 1.7 & $-24.6$ & $-19.4$ & $K$ band \\
\\
$G1$ & 7.4 & $-21.1$ & $-14.5$ & $B$ band \\ 
$G1$ & 3.7 & $-23.8$ & $-17.3$ & $K$ band\\
\end{tabular}
\end{center}
\caption{\label{magz0} Mass-luminosity ratios and characteristic
magnitudes in the $B$ and $K$ bands for the $G0$ and $G1$ simulations
at $z=0$.}
\end{table}

\begin{table}
\begin{center}
\begin{tabular}{c c c c c c}
Simulation & $\mathcal{M}_*$ & $\mathcal{M}_\mathrm t$  \\
\hline
\\
$G0$ ($z=1$) & $-21.4$ & $-16.8$ & $B$ band \\
$G0$ ($z=1$) & $-24.2$ & $-19.6$ & $K$ band \\
\\
$G1$ ($z=1$) & $-20.3$ & $-13.9$ & $B$ band \\ 
$G1$ ($z=1$) & $-23.0$ & $-16.7$ & $K$ band \\
\end{tabular}
\end{center}
\caption{\label{magz1} Characteristic magnitudes in the $B$ and $K$
bands for the $G0$ and $G1$ simulations at $z=1$ estimated with the mass-to-light ratios at $z=0$ given in Table~\ref{magz0}.}
\end{table}

\begin{figure}
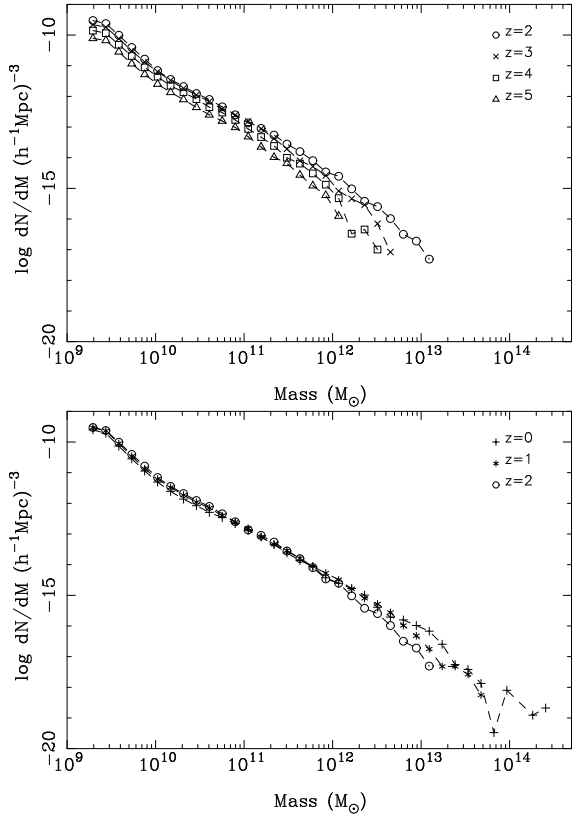

\begin{center}
\includegraphics[height=7.5cm, angle=-90]{fig7a.ps}
\includegraphics[height=7.5cm, angle=-90]{fig7b.ps}
\caption{Dark matter halo mass function at different redshifts for the $G0$ simulation.}
\label{fctmHL_L32}
\end{center}
\end{figure}

\begin{figure}
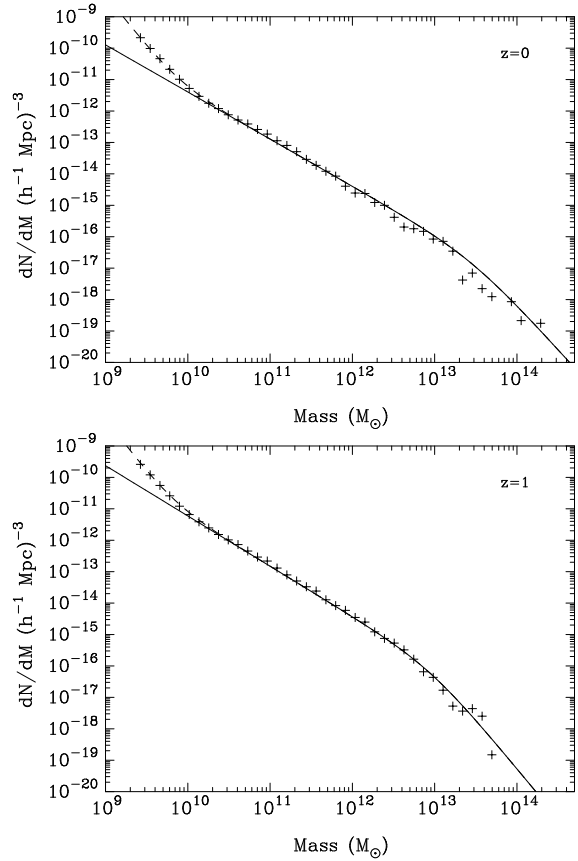

\begin{center}
\includegraphics[height=7.5cm, angle=-90]{fig8a.ps}
\hspace{.5cm}
\includegraphics[height=7.5cm, angle=-90]{fig8b.ps}
\caption{Analytical fits (solid curve) of the dark matter halo mass
  function (cross) for the $G0$ simulation at $z=0$ and $z=1$,
  computed using Eq.~(\ref{fit2pl}). Dashed curve (identical to the solid
  curve at $M>10^{10}$ $M_{\odot}$) is a three power-law analytical
  fit.}
\label{fctmHL_fit}
\end{center}
\end{figure}

From these mass-to-light ratios we estimate the characteristic
magnitudes $\mathcal{M}^*$ and $\mathcal{M}_\mathrm t$ corresponding
to the characteristic masses $M_*$ and $M_t$ (in the expression
$\mathcal{M} - \mathcal{M}_{\odot} = -2.5 \times \dt{log}
(L/L_{\odot})$ solar magnitude values are taken to be
$\mathcal{M}_{b_J}^{\odot}=5.3$ and $\mathcal{M}_{K}^{\odot}=3.3$).
Results are given in Table \ref{magz0}. In the $B$ band
$\mathcal{M}_*$ is brighter than observational values (between -20 and
-20.5, see Table~\ref{litt}).  In the $K$ band $\mathcal{M}_*$ is in
agreement with observations (around -24). Indeed the infrared band is
a better tracer of the stellar content, and is not dominated by young
stellar population strongly emitting in ultraviolet and affected by
dust extinction. Converting mass into luminosity using mass-to-light
ratios is then expected to be more reliable in the $K$ band.  The
transitional magnitudes defined in \cite{Loveday97} and \cite{Zucca}
are $-14.8$ and $-17.7$ (Table~\ref{litt}), respectively, and our
transitional magnitudes in Table~\ref{magz0} are consistent with these
values. Making the crude assumption that the mass-to-light ratio is
constant between $z=0$ and $z=1$, we derive the characteristic
magnitudes at $z=1$ corresponding to the characteristic masses quoted
in Table \ref{fitGG}. Table~\ref{magz1} shows that $\mathcal{M}_*$ is
roughly half a magnitude fainter in the $B$ band than at $z=0$ and
this trend is similar to the evolution with redshift of $L_*$ in the
survey Autofib \citep{Ellis96}.
\\

Before going further into the description of the galaxy population we
compare the galaxy mass function with the dark matter halo mass
function. Figure~\ref{fctmHL_L32} shows that the mass range extends
from $10^9$ to more than $10^{14} \ \mathrm M_{\odot}$. At any
redshift the number density of dark matter halos per bin of mass
increases as the mass decreases.  No strong decrease at the high-mass
end is seen, as in the galaxy mass function, at low redshift because
of our small computational volume, this behavior being expected at
much higher mass \citep{Jenkins2001}. A slight decrease is seen around
$10^{13} \ \mathrm M_{\odot}$ and we thus adopt a two power-law fit in
the range $M>10^9 \ \mathrm M_{\odot}$, such as:
\begin{equation}
\label{fit2pl}
\frac{\ud N}{\ud M} =
\frac{\phi_*}{M_{\mathrm t_1}} \left( \frac{M}{M_{\mathrm t_1}} \right)^{\alpha} \left[ 1+ \left(\frac{M}{M_{\mathrm t_1}}\right)^{\gamma} \right]^{-1}
\end{equation}
where $\phi^*$ the normalization parameter, $\alpha$ and $\gamma $ the
slopes in the high and low-mass range, respectively, and $M_{\mathrm
  t_1}$ a transitional mass between the two power-laws. Fits and their
parameters at $z=1$ and $z=0$ are described in Fig.~\ref{fctmHL_fit}
(solid curve) and Table~\ref{fitHL}. The $M_{\mathrm t_1}$ value is
$3.4 \times 10^{13} \ \mathrm M_{\odot}$ at $z=0$ and around
$9.10^{12}$ at $z=1$. The halo mass functions are much steeper than
the galaxy mass functions in the intermediate mass range. 

However the halo mass functions steepens at the low-mass end,
$M<10^{10} \ \mathrm M_{\odot}$, and the analytical function in
Eq.~(\ref{fit2pl}) does not correctly fit the mass functions for the
entire mass range. We then use a three power-law function by
multiplying $\ud N/\ud M$ by the term $[1+ (M/M_\mathrm t)^{\beta}]$,
similarly to the galaxy mass function.  The fit is shown in
Fig.~\ref{fctmHL_fit} by the dashed curve and Table~\ref{fitHL}
gives this second set of now 6 parameters.

\begin{table*}
\begin{center}
\begin{tabular}{l l l l l l l}
& $\phi_*$ & $M_{\mathrm t_1}$ & $\alpha$ & $\gamma$ & $M_\mathrm t$ & $\beta$ \\
\hline
\\
$z=0$ & 0.000426 & $3.40 \times 10^{13}$ & $-1.60$ & 1.39  \\
      & 0.000743 & $2.95 \times 10^{13}$ & $-1.50$ & 1.40 & $7.13 \times 10^9$ & $-1.86$ \\
$z=1$ & 0.001 & $9.02 \times 10^{12}$ & $-1.62$ & 1.57 \\
      & 0.001 & $9.02 \times 10^{12}$ & $-1.60$ & 1.57 & $6.11 \times 10^9$ & $-1.89$ \\
\end{tabular}
\end{center}
\caption{\label{fitHL} Fit parameters
(Eq.~(\ref{fit2pl})) computed for the dark matter halo mass function
(Fig.~\ref{fctmHL_fit}) for the $G0$
simulation at $z=0$ and $z=1$. The characteristic masses are expressed in $\mathrm M_{\odot}$ and the normalization parameter in $h^3 \mathrm{Mpc}^{-3}$.}
\end{table*}

\subsection{Epoch of formation}

The galaxy mass function characterizes the galaxy population at a
given epoch and does not give any information about the background of
objects. Catalogs are likely to mix galaxies with different
properties. The epoch of formation is then a first insight into their
history. Since each stellar particle carries a formation epoch $a_*$
and a mass $m_*$, the epoch of formation of any object is determined
from each formation epoch of its stellar particles weighted by their
mass, $\sum (a_*m_*)/\sum m_*$.  

\begin{figure}
\begin{center}
\includegraphics[height=8cm, angle=-90]{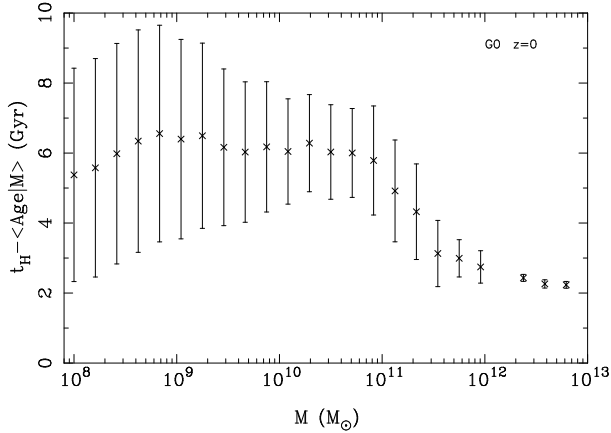}
\caption{Conditional mean of the galaxy-like object epoch of formation given a mass $M$, $\langle t|M \rangle$, for the $G0$ simulation catalog at $z=0$ (today is at the upper end of the ordinate axis). The dispersion $\pm \sigma$ around the conditional mean is shown by the error bars.}
\label{age}
\end{center}
\end{figure}

\begin{figure}
\begin{center}
  \includegraphics[height=8.5cm, angle=-90]{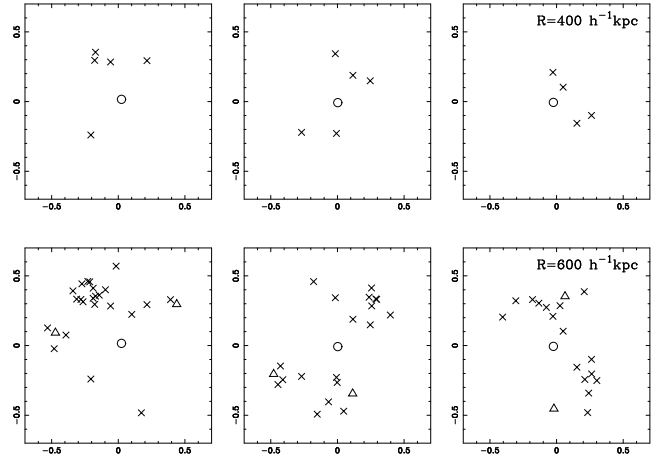}
\caption{Top panel : Projected distribution of the galaxy-like objects inside a radius $<R=400 \ h^{-1}\ \mathrm{kpc}$ around the mass center of three dark matter halos with a mass higher than $10^{13} \ \mathrm M_{\odot}$, for the $G0$ simulation. The halo mass centers are at the origin of each plots, scaled in $h^{-1}\ \mathrm{Mpc}$. Each symbol represents the mass range of galaxies: $10^8<M<10^9 \ \mathrm M_{\odot}$ (cross), $10^9<M<10^{11}\ \mathrm M_{\odot}$ (triangle), $M>10^{11} \ \mathrm M_{\odot}$ (circle). Lower panel: Same as the top panel, but for $R=600 \ h^{-1}\ \mathrm{kpc}$.}
\label{HL_gal}
\end{center}
\end{figure}

Figure~\ref{age} displays the conditional mean of the formation epoch
of galaxies given a mass $M$, $\langle t|M \rangle$ (the age is the
Hubble time minus the epoch of formation), computed for the catalog at
$z=0$. The dispersion around the mean is shown by the error bars. This
quantity is only shown for the $G0$ simulation as the conclusions are
the same for $G1$. The epoch of formation decreases with increasing
mass, for $M>10^{11} \ \mathrm M_{\odot}$ galaxies, whereas the
low-mass galaxies have an epoch of formation globally constant.
High-mass objects show early epoch of formations (see also
\cite{Pearce2001}) but low-mass objects have formed more recently.
The low dispersion around the conditional mean for the high-mass
galaxies reflect the fact that their star formation has considerably
slowed down at low redshift, contrary to the low-mass galaxies showing
a large dispersion. This suggests that a part of the low-mass objects
underwent very recently, or currently have, a star formation activity
at low redshift. Another part of these objects have formed, like the
high-mass objects, at high redshift and have stopped their star
formation. Our results show that most of the stellar mass have formed
by redshift $z=1$, 68$\%$ in the $G0$ simulation (37$\%$ by redshift
$z=2$ and 16$\%$ by redshift $z=3$). In the $G1$ simulation 61$\%$
have formed by redshift $z=1$ and 12$\%$ by redshift $z=3$. This trend
and the fact that the most massive systems have quite old epochs of
formation are similar a part of \cite{Springel} conclusions, although
they have conducted a sophisticated study of the evolution of the star
formation rate, using a large set of high-resolution simulations based
on SPH methods and including star formation, supernova feedback and
galactic outflows. Therefore the galaxy catalog seems to be populated
on the one hand by high-mass galaxies being in majority early-formed
galaxies (see Fig.~\ref{fctmGG_fit} and Fig.~\ref{age}). On the other
hand, the low-mass end of the galaxy mass function is dominated either
by early-formed galaxies or late-formed galaxies with a star formation
activity.  Separating these two last sub-populations would change the
shape of the galaxy mass function. Regarding the fact that the faint
end of the galaxy luminosity function is likely to be populated by
star-forming galaxies \citep{Zucca, Madgwick2002}, this
numerical result is quite encouraging.  \\

The population of high-mass, early-formed galaxies could be the
observational counterpart of red, passive elliptical galaxies in the
center of galaxy clusters. To address this issue we focus on the
galaxy-like object distribution inside the highest mass dark matter
halos, with $M>10^{13} \ \mathrm M_{\odot}$.  Figure~\ref{HL_gal}
displays the galaxy distribution at a distance $<R$ around the mass
center of three dark matter halos, randomly chosen in the
computational volume among the most massive ones.  Galaxies are
plotted, with symbols according to their mass range, inside a radius
of $R=400$ (top panel) or $600 \ h^{-1}$ kpc (lower panel) around each
mass center halos. Dark matter halos host a whole population of
galaxies, from $M=10^8$ to more than $10^{11} \ \mathrm M_{\odot}$.

It is remarkable that each high-mass halo includes in its center a
high-mass galaxy-like object (whose position at center is not defined
a priori). Moreover the biggest halo contains the biggest galaxy-like
object of the catalog. It has been pointed out that the most massive
galaxy-like objects are also old objects and this result is consistent
with the observational evidence that galaxy clusters have cD type
galaxy in their center. It is interesting to note that inside a radius
of $R=400\ h^{-1}$ kpc no galaxy-like object with intermediate mass,
between $10^9$--$10^{11} \ \mathrm M_{\odot}$, is found in the
proximity of the high-mass galaxy, although low-mass objects, $M<10^9
\ \mathrm M_{\odot}$, are present.  Galaxies of intermediate mass
appear when the radius around the mass center increases (lower panel).

\begin{figure}
\begin{center}
\includegraphics[height=8cm, angle=-90]{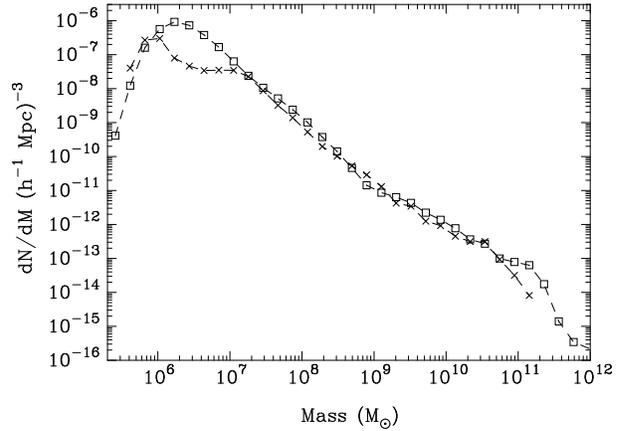}
\caption{Galaxy-like object mass function for the $GP$ (cross) and $G1$ (square) simulations at $z=0$.}
\label{fctmGG_phot}
\end{center}
\end{figure}

\begin{figure}
\begin{center}
\includegraphics[height=8cm, angle=-90]{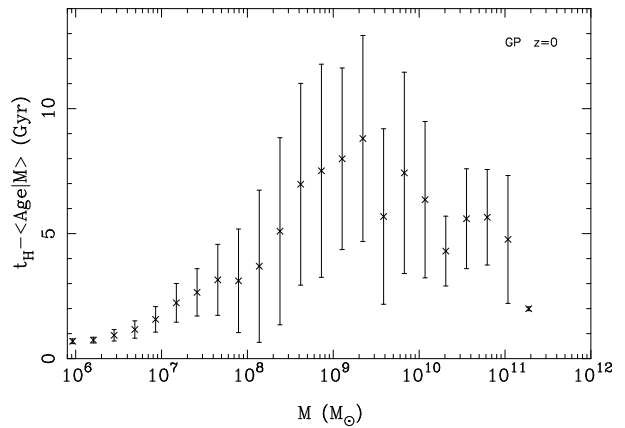}
\caption{Same as Fig.~\ref{age} but for the $GP$ simulation.}
\label{age_phot}
\end{center}
\end{figure}

\section{Photoionization processes}
\label{phot}

We now turn to the $GP$ simulation including an ultraviolet background
radiation (see Section~\ref{simul}) to analyze the influence of
photoionization processes on the galaxy-like object properties. The
same properties as in the previous section are discussed. Figure
\ref{sfr} overplots the cosmic star formation rate density for the
$GP$ simulation.  Although the amplitude is lower than for $G1$, the
general trend is not affected by photoionization processes and the SFR
density decreases from $z \sim 2$ to the present. Recall that the $GP$
simulation has a smaller computational box length than $G0$ and $G1$
and that non-linear long wavelengths are likely to be missing at low
redshift. A dip appears around $z=6$ \citep{Barkana2000} suggesting
that low-mass stellar particles do not form above this redshift in
low-mass structures where gas is now heated by photoionization
processes.  Indeed the gas distribution in the temperature-density
diagram (middle panel in Fig.~\ref{iso}) shows that these processes
are dominant over cooling processes only in low and middle density
regions.

Figure \ref{fctmGG_phot} compares the galaxy mass function for the
$GP$ and $G1$ simulations at $z=0$. It reveals the influence of
photoionization in low-density regions: the most dramatic differences
are seen at the low-mass end, $M<10^7 \ \mathrm M_{\odot}$. The slopes
and characteristic masses are similar as for $G1$. The slight decrease
at the high-mass end is likely due to the smaller computational
volume. The formation epochs of galaxies are plotted in
Fig.~\ref{age_phot} down to $10^6\ \mathrm M_{\odot}$, below this mass
the galaxy-mass function decreases. The plot presents much more
dispersion at $M>10^8\ \mathrm M_{\odot}$ than in Fig.~\ref{age}, the
conditional mean formation epoch varying between 4 and 9 Gyr instead
around 6 in the $G0$ simulation. Galaxies with a mass lower than
$10^8\ \mathrm M_{\odot}$ form at early epochs since the formation of
low-mass stellar particle stops at low redshift (see
Fig.~\ref{fctmGL}).  Similarly to previous results (see
Fig.~\ref{age}) the highest mass galaxies seem to form at early
epochs. Our results suggest that, with our choice for the $F(z)$
function, the photoionization processes have no dramatic influence on
galaxy formation over a mass range higher than $10^7\ \mathrm
M_{\odot}$ (see also \cite{Quinn,Weinberg97}).  More numerical
investigations are nevertheless required to allow definitive
conclusions.

\section{Non-equipartition processes}
\label{s3t}

We have investigated in Paper I the influence of additional
dissipative processes on the cosmological plasma: non-equipartition
processes between ions, neutrals and electrons may change the gas
thermodynamic properties. The astrophysical implications are now
discussed and quantified.

\begin{figure}
\begin{center}
\includegraphics[height=7.5cm, angle=-90]{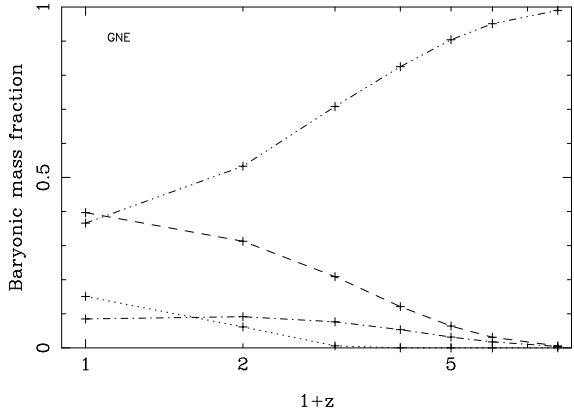}
\caption{Same as Fig.~\ref{fracL32}, but for the $GNE$ simulation without galaxy formation.}
\label{fracL16_s3t}
\end{center}
\end{figure}

Paper I shows that the low-density, outer regions of gravitationally
bound structures are found to be warmer in simulations including
non-equipartition processes than in simulations in which equipartition
is forced. This results in a warm gas fraction, illustrated by the gas
distribution in Fig.~\ref{iso} (bottom panel).  Moreover we show that
the cooling timescale of the warm plasma is longer than the cooling
timescale of the same regions in simulations with forced
equipartition. Figure~\ref{fracL16_s3t} gives the baryonic mass
fractions in different ranges of temperature for the simulation
including non-equipartition processes from Paper I, the $GNE$
simulation but without galaxy formation, and should be compared with
Fig~\ref{fracL16} for $G1$: the fraction of gas with a temperature
higher than $9.10^3$ K is similar in both figures but the gas is not
distributed in the same phases: at $z=5$, for example, 3.1$\%$ of the
plasma constitutes the warm phase and 6.4$\%$ the cold phase in the
simulation with non-equipartition processes, whereas these phases are
0.2$\%$ and 9.1$\%$ in $G1$, respectively.  The decrease in available
cold gas needed for a galaxy to form is expected to affect the galaxy
formation process.

\begin{figure}
\begin{center}
\includegraphics[height=6.5cm]{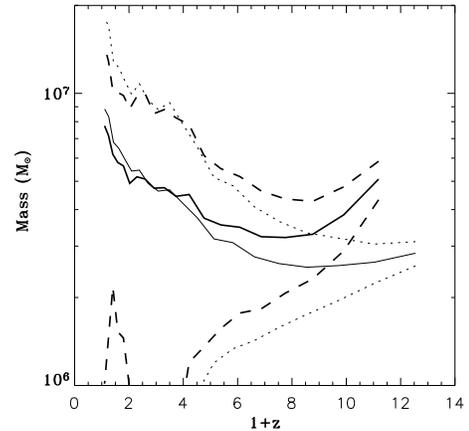}
\caption{Evolution with redshift of the stellar particle mean mass (solid lines) with the dispersion $\pm \sigma$ around the mean (dashed and dotted lines) for the $GNE$ (thick lines) and $G1$ simulations (thin lines).}
\label{mstar_s3t}
\end{center}
\end{figure}

We then analyze the $GNE$ simulation, the same simulation as performed
in Paper I but now including galaxy formation (Table~\ref{param}). As
the influence of non-equipartition processes is likely to be dominant
at epochs before the end of the reionization epoch, we do not include
photoionization processes and discuss the results at redshifts lower
than 10 to mimic what happened before this epoch.

\begin{figure}
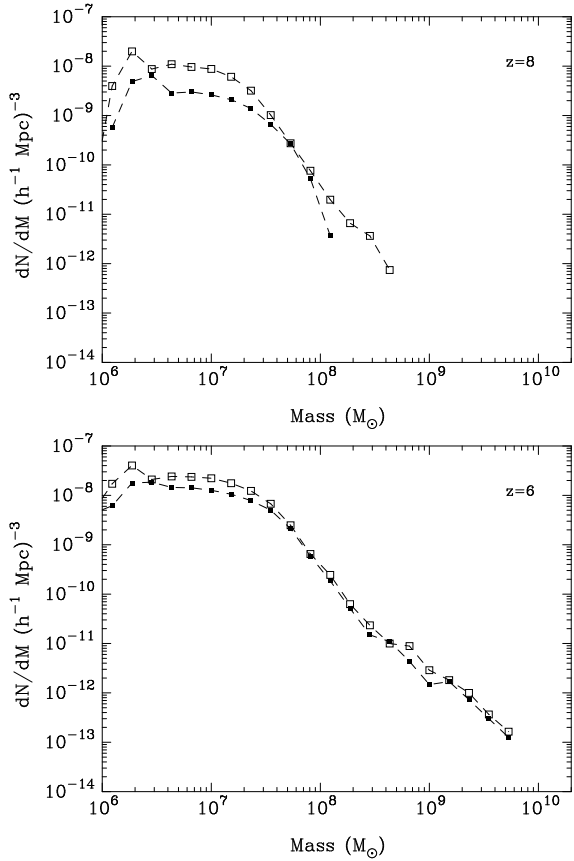

\begin{center}
\includegraphics[height=7.5cm, angle=-90]{fig14a.ps}
\includegraphics[height=7.5cm, angle=-90]{fig14b.ps}
\caption{Galaxy-like object mass function for the 
$GNE$ (square) and $G1$ (open square) simulations at $z=8$ and $z=6$.}
\label{fctmGG_s3t}
\end{center}
\end{figure}

Figure \ref{mstar_s3t} compares the evolution with redshift of the
stellar particle mean mass between the $GNE$ and $G1$ simulations. The
galaxy formation process starts at a lower redshift in the former
simulation implying a higher mean mass.  Such differences are also
seen in the stellar particle mass functions (bottom panel in Fig.
\ref{fctmGL}). Figure~\ref{fctmGG_s3t} compares the galaxy-like object
mass function in $GNE$ and $G1$.  Differences can be seen at the low
and high-mass ends: at $z=8$ the numerical density of galaxies with a
mass higher than $5.10^6 \ \mathrm M_{\odot}$ is $0.2 \ h^3
\mathrm{Mpc}^{-3}$ in the $G1$ simulation but only $0.08 \ h^3
\mathrm{Mpc}^{-3}$ in $GNE$. This represents a 60$\%$ decrease in the
number of objects in the latter simulation.  Moreover the mass of the
biggest galaxies is lower in $GNE$, $10^8$ against $5.10^8 \ \mathrm
M_{\odot}$. At $z=6$ differences are less dramatic and only appear for
the low-mass end of the galaxy mass function. As discussed in Paper I
the influence of non-equipartition processes is dominant in shallow
potential wells making gravitational compression unable to heat
cosmological plasma to temperatures higher than $10^6$ K. Added to the
fact that potential wells become deeper as cosmological evolution
proceeds, this explains the decrease in the fraction of out of
equilibrium plasma at lower redshift.

The effects on galaxy properties due to thermodynamic modifications of
the inter-galactic medium are clearly shown here. A follow-up paper
(Courty $\&$ Alimi, in preparation) using the same simulations as in
the present paper will discuss changes in the clustering properties of
the galaxy-like objects, giving some insights into the physical origin
of cosmological bias.

\section{Conclusions}
\label{conclu}

We examine global galaxy properties and the connection between galaxy
formation and the thermodynamics of the cosmological gas in Eulerian
hydrodynamical simulations. The dominant processes known to play a
role in galaxy formation are included: gravitation, shock heating and
cooling processes. The galaxy formation model consists of the
identification in the gas distribution of dense and cold regions. A
part of this gas is turned into a stellar particle whose collection
provides a catalog of galaxy-like objects.  In addition, the
identification of dense regions in the dark matter distribution
provides a dark matter halo catalog. We estimate a number of
properties, the cosmic star formation rate density, the galaxy-like
object/dark matter halo mass function and the formation epoch of
galaxies.

The most striking result is that considering in the simulations these
dominant processes gives galaxy properties consistent with
observations. The cosmic star formation rate density shows a peak
around $z \sim 3$ and reproduces at low redshift the evolution of the
observational star formation rate density, namely the sharp decline
between $z=0$ and $z=1$. More than 60$\%$ of the stellar mass has
formed by redshift $z=1$. The galaxy-like object mass function
presents a significant population of low-mass galaxies and shows an
evolution with redshift.  Moreover, the galaxy mass function is well
described at low redshift in the range $M>10^8 \ \mathrm M_{\odot}$ by
an analytical function combining a standard Schechter function with a
$\beta$ power-law at the low-mass end. The fit, parameterized by two
characteristic masses and two slopes, appears to be in good agreement
with the observed galaxy luminosity function. The dark matter halo
mass function is found to be well fitted by a three power-law function
in the range $10^9<M<10^{13} \ \mathrm M_{\odot}$ and is steeper than
the galaxy mass function in the intermediate mass range. The estimate
of the galaxy formation epoch shows that high-mass galaxies form, in
the majority, at early epochs.  Moreover the galaxy distribution
around the mass center of the highest mass dark matter halos shows
that these halos include in their center a high-mass, old galaxy-like
object. On the other hand the low-mass galaxies present a large
dispersion around their conditional mean formation epoch, suggesting
that they include recently formed stellar material.

The galaxy formation model is simple enough for the galaxy formation
to depend on the thermodynamic properties of the baryonic matter and
its distribution. Introducing the photoionization processes has a
dominant effect in the low-density regions.  This turns into a
decrease in the density of the lowest mass objects, $M<10^7 \ \mathrm
M_{\odot}$, but no drastic change is seen in the galaxy mass function
at the high-mass end, neither in the galaxy formation epoch at
intermediate and high mass end, although this quantity shows more
dispersion. However these results could depend on the adopted
intensity of the ultraviolet background radiation. On the other hand
the introduction of the non-equipartition processes between the
electrons, ions and neutrals of the cosmological plasma results in a
warmer plasma at high redshift and in not too dense regions, than in
simulations in which equipartition between species is forced. Hence
the longer cooling timescale delays star formation. The galaxy-like
object mass function then shows a decrease in the density of objects
at the low-mass end.

We also compare two simulations with different computational box
lengths and show that galaxy properties depend on the resolution. The
amplitudes of the star formation rate density and the galaxy mass
function are higher for the simulation with the highest resolution
than with the lowest. Indeed adopting a different resolution affects
the history of the accretion and gas cooling inside the dark matter
potential wells. This results in a change in the fraction of available
cold gas at a given time, modifying therefore the galaxy-like object
population. We have chosen middle-size box lengths and standard
resolutions to allow the formation of high as well as low-mass
objects. This work should then be seen as a qualitative analysis of
the galaxy formation process and no calibration on observations at
$z=0$ is adopted. Altogether the galaxy properties draw a consistent
picture of the galaxy formation process and some common features are
independent of the resolution: the evolution of the cosmic star
formation rate density, the shape of the galaxy-like object mass
function, the facts that high-mass galaxies form at early epochs,
residing in the highest mass dark matter halos, and that low-mass
galaxies, for some of them, form at later epochs, thus showing a star
formation activity.

\bibliography{biblio}

\begin{acknowledgements}
  
  Numerical simulations of this paper were performed on NEC-SX5 at the
  I.D.R.I.S. computing center (France). SC is grateful to A. Hopkins
  for having kindly provided the observational data of the star
  formation rate density and to Gunnlaugur Bj\"ornsson for a careful
  reading of the paper. SC acknowledges support from a special grant
  from the Icelandic Research Council. We thank our referee, Naoki
  Yoshida, for his helpful comments on the manuscript.

\end{acknowledgements}

\appendix

\section{The stellar particle mass function}
\label{A1}

\begin{figure}
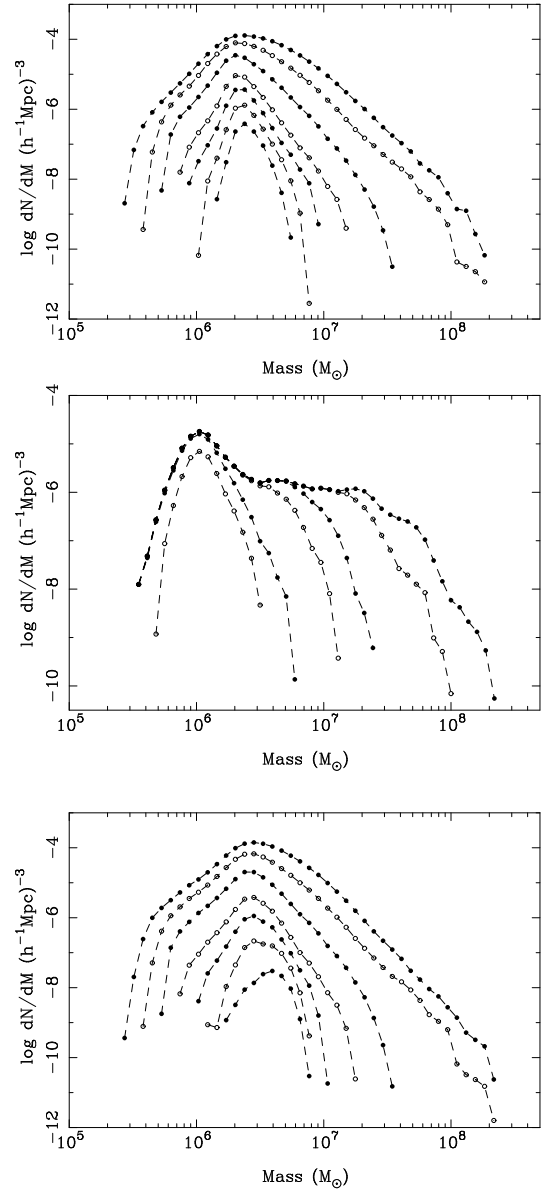

\centering
\includegraphics[height=7cm, angle=-90]{figA1a.ps}
\vspace{.5cm}
\includegraphics[height=7cm, angle=-90]{figA1b.ps}
\vspace{.5cm}
\includegraphics[height=7cm, angle=-90]{figA1c.ps}
\caption{Evolution with redshift of the stellar particle mass
  function for the $G1$, $GP$ and $GNE$ simulations (from top to bottom
  panels, respectively). The mass function is shown at $z=$9, 8, 7, 6,
  4, 2, 0 for the $G1$ and $GNE$ simulations and at $z=$8, 7, 5, 4, 2,
  0 for the $GP$ simulation (alternatively filled and open circles,
  from bottom to top).}
\label{fctmGL}
\end{figure}

\begin{figure}
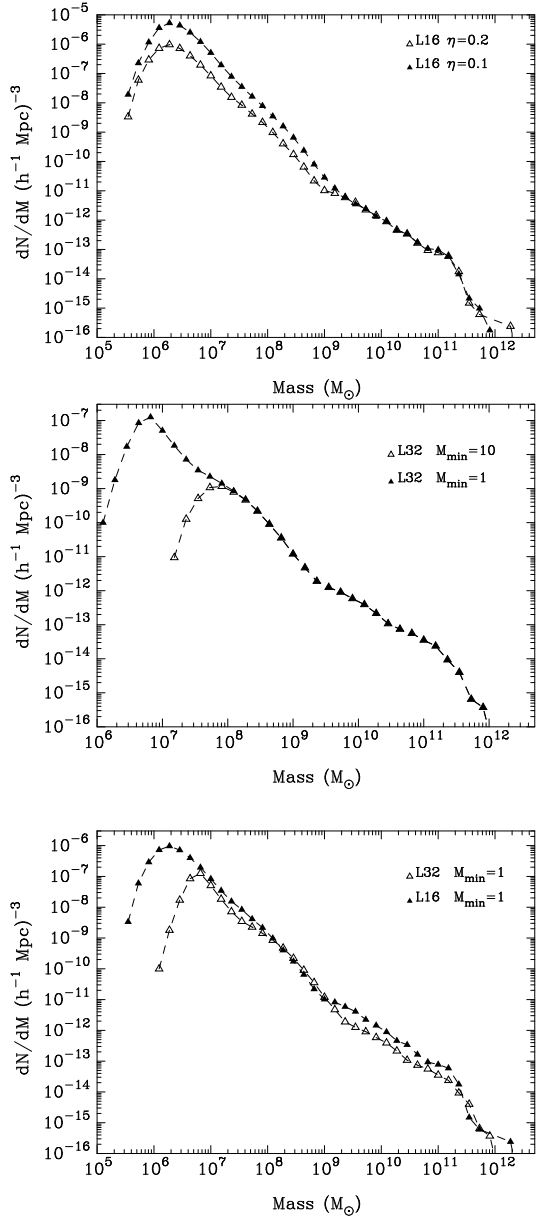

\begin{center}
\includegraphics[height=7cm, angle=-90]{figB1a.ps}
\vspace{.5cm}
\includegraphics[height=7cm, angle=-90]{figB1b.ps}
\vspace{.5cm}
\includegraphics[height=7cm, angle=-90]{figB1c.ps}
\caption{Galaxy-like object mass function for the $G0$ ($L_\mathrm{box}=32$) and $G1$ ($L_\mathrm{box}=16$) simulations at $z=0$ computing by considering different values of $\eta$ (top panel) and $M_\mathrm{min}$ (middle panel). The bottom panel compares both resolutions.}
\label{fctm_comp}
\end{center}
\end{figure}

Figure~\ref{fctmGL} presents the evolution with redshift of the
stellar particle mass function in the three kinds of simulations.  In
$G1$ the stellar particle mass ranges between $3.10^5$ and $2.10^8 \ 
\mathrm M_{\odot}$ at $z=0$, typically that of the globular clusters.
We note that the mass range of the particles increases as the redshift
decreases: at any epoch, not only have low-mass stellar particles
formed, but also larger ones.

In the $GP$ simulation (middle panel) the formation of the low-mass
stellar particles is stopped when photoionization processes become
dominant, for redshifts less than 7 (after the epoch of the sharp
steepness of the ultraviolet background radiation spectrum,
Fig.~\ref{specphot}). At lower redshift the mass function presents an
evolution for particles with mass higher than $3.10^6 \ \mathrm
M_{\odot}$ and the general shape seen in the top panel is modified.

The bottom panel displays the mass functions for the simulation with
non-equipartition processes: at high redshift the general shape of
upside down ``V'' seen for the $G1$ simulation, with a similar maximum
(around $2.10^6 \ \mathrm M_{\odot}$ in $G1$) shows that the numerical
density of stellar particles is much lower in the $GNE$ simulation
than in $G1$.

\section{Sensitivity to parameters}
\label{sensitivity}

We briefly discuss how the galaxy mass function is sensitive to the
parameters involved in determining the galaxy-like objects and to the
resolution.

The top panel of Fig.~\ref{fctm_comp} shows the influence of the link
parameter $\eta$ used to group stellar particles and initially
fixed to 0.2 (see Section~\ref{simul}). Taking $\eta=0.1$ results in a
steeper $\beta$ slope. The catalog then includes a larger number of
low-mass objects. On the other hand, choosing $\eta=0.3$ would result
in a shallower $\beta$ slope. Nevertheless there is no modification in
the $\alpha$ power-law range.

The middle panel of Fig.~\ref{fctm_comp} compares the galaxy mass
function at $z=0$ in the $G0$ simulation considering
$M_\mathrm{min}=1$ and $M_\mathrm{min}=10$. The parameter
$M_\mathrm{min}=1$ means that even a single stellar particle is
identified as a galaxy-like object. This extends the mass function
towards lower mass, shifting the decrease at the low mass end, but
does not change the rest of the mass function.

Finally the bottom panel of Fig.~\ref{fctm_comp} illustrates the
influence of the resolution. As already seen with the fits in
Fig.~\ref{fctmGG_fit} the $\alpha$ and $\beta$ slopes and the
characteristic mass $M_*$ are roughly similar for the two simulations.
But in the high-resolution simulation, $G1$, the transitional mass
$M_\mathrm t$ and the decrease at the low mass end are shifted towards
lower mass, this simulation being able to form lower mass objects.

\end{document}